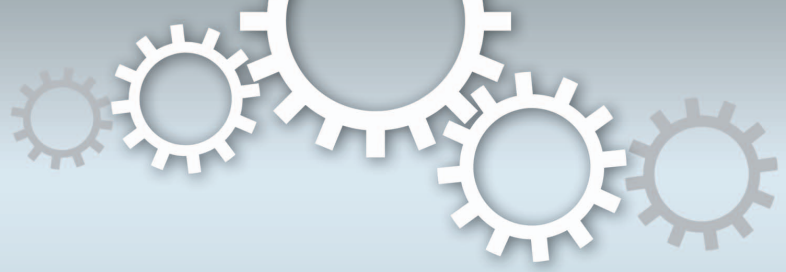

**OPEN**

# GPS source solution of the 2004 Parkfield earthquake

N. Houlié[1,2,3], D. Dreger[1] & Kim, A.[4,5]

[1]Berkeley Seismological Laboratory - University of California, [2]ETH, SEG, Zurich, CH, [3]ETH, GGL, Zurich, CH, [4]Schlumberger, Cambridge, UK, [5]Graduate School of Nanobioscience, Yokohama City University, 22-2 Seto, Kanazawa-ku, Yokohama 236-0027, Japan.

We compute a series of finite-source parameter inversions of the fault rupture of the 2004 Parkfield earthquake based on 1 Hz GPS records only. We confirm that some of the co-seismic slip at shallow depth (<5 km) constrained by InSAR data processing results from early post-seismic deformation. We also show 1) that if located very close to the rupture, a GPS receiver can saturate while it remains possible to estimate the ground velocity (~1.2 $m/s$) near the fault, 2) that GPS waveforms inversions constrain that the slip distribution at depth even when GPS monuments are not located directly above the ruptured areas and 3) the slip distribution at depth from our best models agree with that recovered from strong motion data. The 95$^{th}$ percentile of the slip amplitudes for rupture velocities ranging from 2 to 5 km/s is ~55 ± 6 cm.

Today, few instruments are able to constrain simultaneously the static and transient displacements for periods between 1 s to 1000 s or more. GPS is one of them. GPS is traditionally known to accurately measure distances and changes of distances through time if network measurements are repeated enabling the detection of surface ground motion including asymmetric deformation due to dislocation at depth[1,2], to constrain material property differences across faults[2–4], or to investigate the relationship between short- and long-term deformation of the continental lithosphere[5–9]. With improved telecommunications links and higher sampling rates, slip distribution models based on GPS data have become an important tool to study processes such as deformation transients from seismic wave propagation or directly from seismic rupture[10–29], changes in ionosphere status[30–35], volcanic plumes[36–38], water vapour delays[39], Early Warning systems[26,40,41] or more recently moment tensor analysis[42–44]. Constrained with GPS data or not, the dislocation amplitude (or fault slip) models occurring at depth during an earthquake are today the foundation of many geophysical studies including seismic hazard[45], rupture modelling[46], post- and inter- seismic deformation[47,48], stress transfer[9,49] or fault topography studies[50]. Meanwhile, the images of the co-seismic slip at depth of a given earthquake seem to depend on the imaging techniques and/or on the type of instrument used. Slip solutions based on static offsets (GPS or InSAR) define the extent of ruptured areas well but potentially incorporate signals from both co- and post-seismic deformation. Inversions of high frequency seismic records offer spatial detail but limits of ruptured fault segment is not well constrained or needs to be defined beforehand. In order to overcome the trade-off between resolution and rupture extent, many studies have tried to combine both static offsets and seismic waveforms in order to minimize the amount of post-seismic deformation to be included in the co-seismic slip model or/and to control the integration of seismic records[51,52–58]. Still, slip distributions vary among authors and processing methodologies[59] (see http://equake-rc.info/ for a extensive collection of fault rupture models for many earthquakes). This can be explained because, with such heterogeneity of instrumentation and sampling rates, the integration of seismic and geodetic data into a common model is not trivial, and the kinematic parameterization of such models is not unique. For example, if source functions are modelled as distributions of double-couples of forces distributed over the rupture area[60] for large events, other assumptions are commonly made in the setup of the inverse problem, such as the rheology and the velocity model, typically 1D, used to generate Green's functions, and inelastic or slow deformation (from rigid block motion to slow slip deformation) cannot be easily constrained by using seismic records only. An ideal setting would allow using both geodetic and seismic datasets as if they were equal in term of quality and use them under the same set of assumptions. To reach this level of confidence, we need to show that consistent slip models at depth can be retrieved using seismic and high-rate geodetic datasets with a same inversion algorithm. The fault system at Parkfield (Figure 1) is a relatively simple, approximately planer, near vertical fault. However, it is interesting to note that the approximate 20% velocity contrast attributed to the contact of the west side granites (faster) with the east side meta-sediments across the SAF[63], whereas for continental geology units maximum differences in elasticity parameters should not reach 30%, approximately 10% in seismic wavespeed, (the maximal value for a oceanic-continental crust contact[3,4,61,62]). The 2004 Parkfield event is interesting because after a year the post-seismic surface deformation is larger than the co-seismic[64,65,84]. For this event, it is therefore critical to distinguish co-seismic and post-seismic deformation to better characterize the co-seismic rupture. We define as "co-seismic" any parameter changing during the first 30 s following the origin time of the event ($T_0$). The post-seismic period is defined as immediately following the co-seismic period. Finally, the density of the monitoring networks and their proximity to the epicenter (60+ seismic and geodetic instruments located within 250 km²) make of the 2004 Parkfield earthquake a very good dataset to test the capability of high-rate GPS in the near-field of large (M6+) earthquakes.

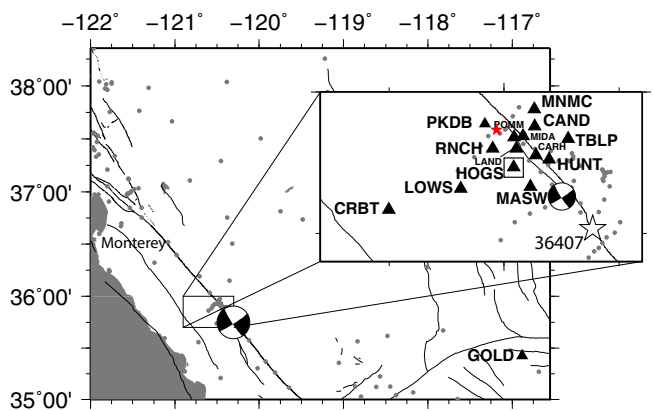

Figure 1 | The GPS and CSMIP strong-motion networks in California and near Parkfield. We locate the CSMIP sensors used in previous studies with small grey circles and GPS receivers with black triangles. The USGS Accelerometer (PHOB), collocated with the GPS site HOGS, is indicated by a white square. We show the location of the 1966 event (red star) and the focal mechanism provided by United States Geological Survey (USGS) for the 2004 Parkfield mainshock. The accelerograph (36407) and GPS site GOLD mentioned in the text are respectively highlighted by a solid-line star and black triangle.

Table 1 | Coseismic displacement at the Parkfield sites. No asymmetry of amplitude can be detected between the two sides of the fault

| | Site | Lat. (deg.) | Lon. (deg.) | De (cm) | Dn (cm) | Dup (cm) |
|---|------|-------------|-------------|---------|---------|----------|
| 1 | CAND | 239.566 | 35.939 | 2.1 | −4.2 | −0.1 |
| 2 | CARH | 239.569 | 35.888 | 1.1 | −1.2 | −0.1 |
| 3 | CRBT | 239.249 | 35.791 | −0.3 | 0.0 | −0.1 |
| 4 | HOGS | 239.521 | 35.866 | −2.2 | 3.5 | −0.9 |
| 5 | LAND | 239.526 | 35.900 | −3.3 | 3.0 | −1.1 |
| 6 | LOWS | 239.406 | 35.829 | −1.1 | 0.2 | −0.2 |
| 7 | MASW | 239.557 | 35.833 | −1.8 | 3.4 | −1.3 |
| 8 | MIDA | 239.541 | 35.921 | 2.1 | −4.5 | −0.9 |
| 9 | MNMC | 239.566 | 35.970 | 0.5 | −3.6 | −0.8 |
| 10 | POMM | 239.521 | 35.919 | −3.6 | 0.6 | −0.5 |
| 11 | PKDB | 239.458 | 35.945 | −3.4 | 0.9 | 0.8 |
| 12 | RNCH | 239.475 | 35.900 | −2.8 | 1.8 | 0.2 |
| 13 | TBLP | 239.639 | 35.917 | 2.6 | −1.8 | 0.9 |
| 14 | HUNT | 239.598 | 35.880 | 3.5 | −3.5 | 0.4 |

At the time of the event, the mini-Plate Boundary Observatory (PBO) GPS network maintained by United States Geological Survey (USGS), University of California at San Diego (UCSD) and University of California at Berkeley (UCB) was composed of 13 receivers recording data at 1 Hz (Figure 1) supplemented by one GPS site sampling data at 30 s (PKDB). The GPS site PKDB is co-located with a very-broadband seismic sensor PKD (STS-2) maintained by Berkeley Seismological Laboratory (BSL) of UCB. The mini-PBO GPS network was deployed around the epicenter of the 1966 Parkfield earthquake. At the time of the event, few GPS sites were located near the southern end of the Parkfield segment, near the location of the epicenter. Each monument was equipped with a ASHTECH Z-12 receiver with a Choke-Ring antenna. One GPS site (HOGS) was co-located with an accelerograph (PHOB) of the California Strong Motion Instrumentation Program (CSMIP) network (Figure 1).

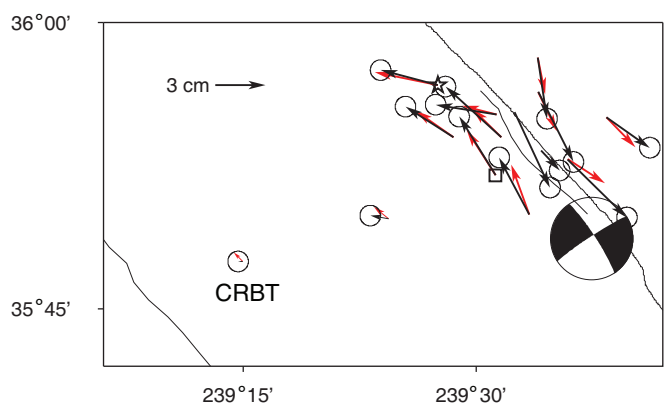

Figure 2 | Observed co-seismic displacements during the mainshock (black). The displacement at the CRBT site is less than 3 millimeters. The displacement observed at PKDB (star) indicates the rupture along the fault is limited to the south. All the displacement values are presented in Table 1. Co-seismic displacements have been inverted (red) using the OKINV software[1]. Minimum depth of the dislocation = 3.8 km, average slip = 49 cm. We indicate the focal mechanism provided by United State Geological Survey (USGS).

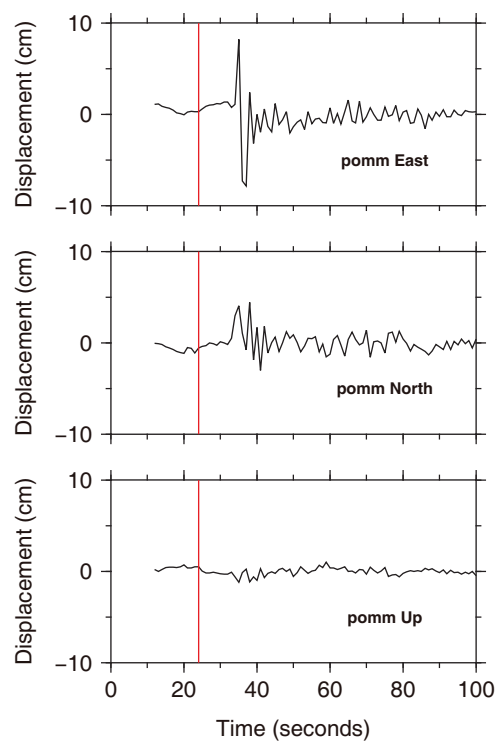

Figure 3 | Ground motion detected at the GPS site POMM with respect to the site CRBT. We indicate the origin time $T_0$ by using a red line.

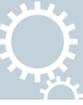

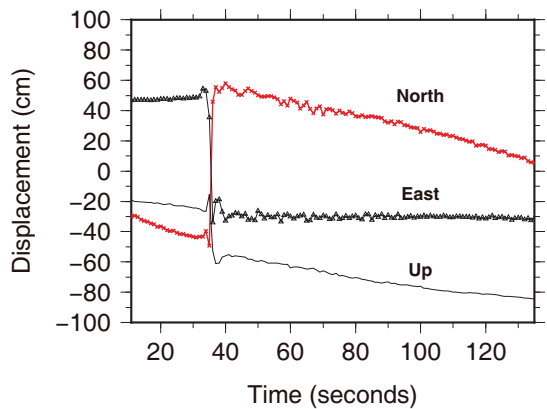

**Figure 4 | Artifact of observation at the GPS site MIDA.** During the rupture, GAMIT processing suggests the peak ground velocity is so high that it resulted in GPS receiver cycle-slip affecting the phase modeling.

## Results

In order to constrain the amplitude of the surface deformation history during the first days following the earthquake as suggested by Langbein et al.[84], we processed the 30s sampling rate data using the GAMIT 10.4 tool suite[66]. The maximum static displacement at the surface is equal to ~5 cm (Figure 2). As expected[67], we find that the displacement of the site GOLD (located 200 km south-west of Parkfield, Figure 1) during the earthquake can be neglected and that the motion of the site CRBT is less than 2 mm (Figure 2). Daily time-series of displacement display a non-linear temporal evolution of the surface deformation during the days and weeks immediately following the mainshock[64,67]. We inverted for the average slip motion at depth for data collected during the first day that followed the earthquake computing the effect at the surface of a buried dislocation in a purely elastic half-space[68] using the nonlinear inversion algorithm OKINV 3.16[1,69]. In order to avoid reaching a local minimum solution, we use 200 Monte-Carlo starts. The inverted solution corresponds to an averaged co-seismic slip at depth of 20 cm. The surface velocity field (Table 1) is shown in Figure 2. The data inversion using 24 h of data implies larger surface displacement than those inferred from the GPS solutions using 20 minutes GPS data[53] suggesting that post-seismic deformation was already on-going during the first 24 h following the earthquake. An inversion of the difference of motion between the two long-term GPS datasets suggests an asymmetry of motion with a fault plane solution shifted to the east with respect to the San Andreas fault (SAF) trace at the surface. This result implies that the elastic response to the rupture during the hours following was asymmetric and/or some post-seismic deformation happened very quickly after the end the earthquake rupture and/or the co-seismic ground motion was asymmetric (the southwestern block moved less than the northern one).

We describe here how the GPS 1 Hz residuals have been processed. This approach has already been used to detect atmosphere disturbances caused by atmospheric volcanic explosions[36,37] or seismic wave propagation in the far field[23]. First, we processed all 1 Hz GPS data available (13 sites) using GAMIT 10.4[66]. We then extracted the LC phase residuals for which the ionosphere contribution has been minimized. At this stage, with phase cycle ambiguities being fixed, clock drifts removed (double differences have been applied) and with troposphere, ionosphere or tide changes being insignificant to the waveform shapes over the first 30 s following the start of the event, phase residuals only highlight the deformation of the network. Each phase residual correspond indeed to the dot product of the ground displacement with the unit vector defined by the Line-Of-Sight (LOS) between a site and a satellite. All single-difference phase residuals available for a given site are inverted in order to retrieve the motion in the local reference frame (East, North and Vertical). Waveforms are 30-seconds long Figure 3 and include 10 seconds of data before the origin time (17:15:24 UTC). To remove any common mode translation of the network, we subtract the motion history of the site CRBT (from now on fixed) to each waveform in the dataset. We compare the GPS displacement waveforms (GPS site HOGS) with acceleration data (converted into displacement) recorded by collocated instruments (PHOB strong motion sensor). In order to compute displacement from acceleration, we corrected the raw accelerograms of the instrument responses and decimated the data to 1 Hz using the SAC2000 software[70]. Filtered strong motion displacement time series exhibit a long-period parabolic drift as expected[71]. In order to remove this artefact of integration, we use a Butterworth filter[72] to high-pass (T = 50 s) the displacement time-series computed from accelerograms. We find that both waveform datasets are in agreement (Figure A2) and conclude that even if GPS and accelerometer instruments have very different designs, inversion of both dataset if sampled at the same frequencies and have the same applied filters should lead to similar models. We focus now on the other waveforms recorded by GPS receivers that are not collocated with seismic instruments. All GPS waveforms final offsets match the expected static offsets except for those computed for the site MIDA. Indeed, the final offset of the GPS site MIDA is close to ~120 cm (Figure 4). We refer to this value as the dynamic offset of MIDA ($d_{dynamics}$). In order to check whether this motion to be related to a software issue, we processed the same data (baseline CRBT-MIDA) by using the software TRACK and did not observe the same processing artefact (Figure 5). To explain such discrepancies, we argue that the rapid carrier phase change at a unique location of the network could not be modelled by using the smooth phase function applied by GAMIT across the GPS network. On the contrary, TRACK's processing strategy, which is based on the phase difference modelling between two sites (assuming one of them is not moving, an assumption that is not true in this case), is more flexible and because ambiguities were fixed properly after the phase slip, the phase change was only an artefact of measurement. We think this artefact is the first GPS clip (or earthquake induced cycle slip) ever observed. As such motion could correspond to a M7+ rupture reaching the surface, unconstrained cycle slips could lead to significant misleading alert for Earthquake Early Warning or detection alert systems[40,41,73] supplemented by the tracking of GPS sites positions. However, we can still use the capability of GPS to clip to calculate the maximum ground motion. Indeed, if we hypothesize that the rapid motion of the site MIDA between two epochs (one second) induced a cycle slip of an unknown number $n$ of carrier cycles and the dynamic offset ($d_{dynamics}$) is almost equal the static offset $d_{static}$ (itself smaller than a phase cycle length), we can retrieve the number of carrier cycle $n$ that slipped during one second by solving the following system of equations:

$$d_{true} \sim d_{dynamics} n \times \lambda_{LC} + \epsilon_1 \quad (1)$$

$$d_{true} \sim d_{static} + \epsilon_2 \quad (2)$$

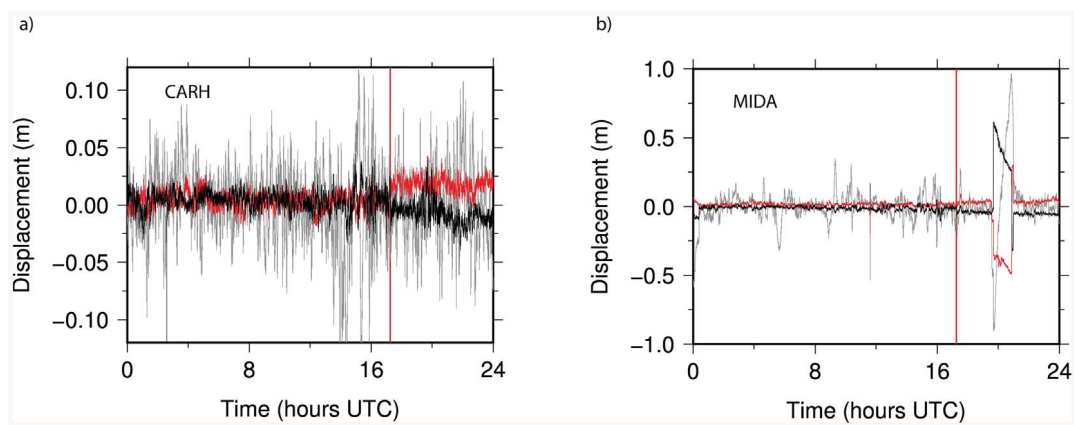

**Figure 5 | Variations of the baseline length components between the sites a) CRBT-CARH and b) CRBT-MIDA.** CRBT is fixed during the processing. East, North and Vertical components are respectively indicated by using red, black and grey lines. The origin time is indicated by a red vertical line. We conclude that the TRACK software could not reproduce the cycle slip shown in Figure 4. Therefore, we exclude the possibility of an internal receiver problem.

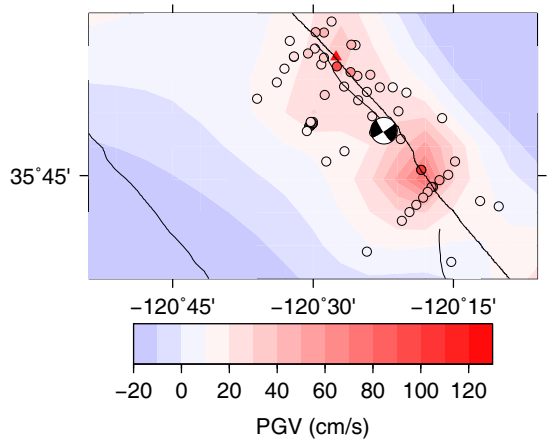

**Figure 6** | **Peak Ground Velocity map compared with the maximum ground velocity measured at MIDA (triangle).** The PGV values obtained from the COSMOS virtual data center (http://db.cosmos-eq.org) have been interpolated using the *surface* function of Generic Mapping Tools on a 5-km interval grid.

Where $\lambda_{LC}$, the wavelength of the ionosphere-free LC carrier equals 24 cm. The true phase change remains unknown because the phase change artefact cannot only be represented by full phase cycles. $\epsilon_1$ represents this uncertainty. We take that $\epsilon_1$ is close to the uncertainty on the GPS carriers phase (2 mm). $\epsilon_2$ represents the post-seismic deformation that occurred during a day, and to be equal or less than $\epsilon_1$. The motion ($d_{static}$) of MIDA is ~5 centimeters (Table 1), we estimate the number of cycles $n$ to be removed of the original time-series to be 5 (~1.2 m). As this motion is completed within a second, we reach the conclusion that the ground velocity at MIDA was close to 1.2 m/s. Such a motion rate is compatible with the Peak Ground Velocity (PGV) observed at the strong-motion seismic site 36407 (located very south of MIDA but within the fault zone at 120.307 W, 35.758 N, 24 km SE from the GPS site MIDA, with PGV = 1.05 m/s). All PGV are presented in Figure 6. Because we do not want to introduce biases in the slip models, we reprocessed the 1 Hz displacement time-series after excluding the site MIDA.

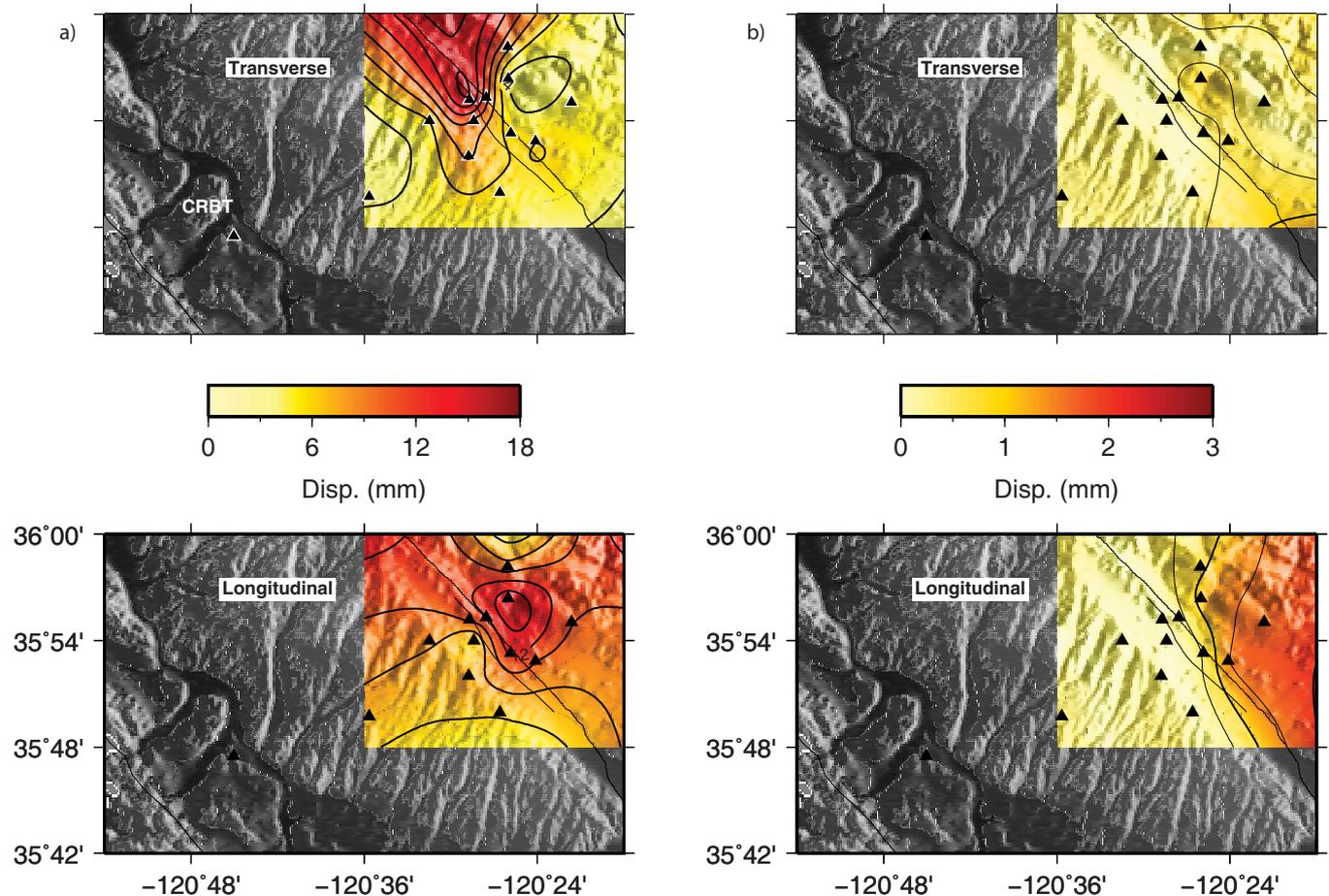

**Figure 7** | **Maximum displacement (a) during the rupture propagation compared to static final offets (b).** We find that both the peak-to-peak amplitudes and the static displacements are ~60% larger on the eastern side of SAF.

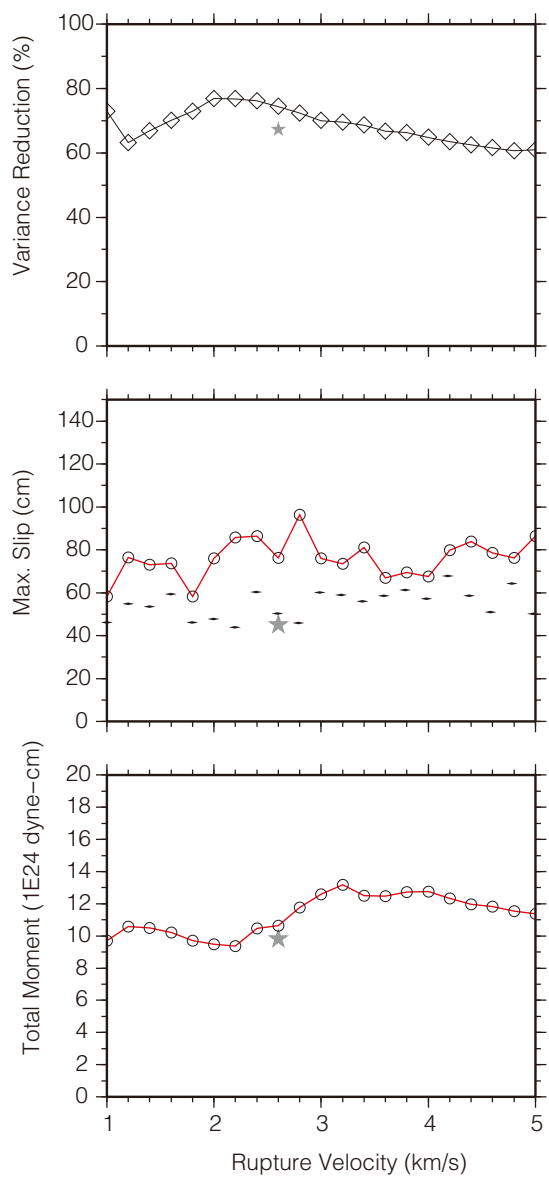

**Figure 8** | (a) Inversion output parameters of the GPS time-series for various rupture velocities (RV) for symmetric velocity models. The best solution proposed by[53], is indicated by a gray star. 95$^{th}$ percentile slip amplitude are indicated by black diamonds. For rupture velocities (RV) between 2.0 and 3.0 km/s, maximum slip is 55 cm.

The 1 Hz GPS time-series show well defined seismic first arrivals and for some sites, nonzero offsets at the end of their displacement history. We find that the motion of the sites located on the eastern side of the SAF is up to 60% larger than the motion of the sites located on the other side of the fault (Figure 7). The asymmetric character of the rupture confirms the observation made with daily displacement time-series. This asymmetry of motion could be due to a limited number of parameters. The first one would be the variation of rheology in the direction perpendicular to the fault. Others could be 1) that the fault plane is not perfectly vertical at depth[50,74] or the velocity model should be different across the fault, which has been tested in previous studies[52,53,75]. It is difficult argue for any one hypothesis. It is clear that the velocity contrast exists from tomographic analysis, and that the fault plane is not perfectly vertical[85]. Thus the third possibility, that there could be anisotropic rheological properties, must be considered only if after accounting for velocity model differences across the fault, 3D structure, and non-vertical fault dip, taking into account the associated uncertainties, still results in a significant residual.

We successfully inverted the GPS waveforms (low-pass filtered at T = 3 s) of 9 of the 13 (CAND, CRBT, HOGS, HUNT, LAND, LOWS, MASW, RNCH, TBLP) high-sampling rate sites using the approach defined by[60,76] for rupture velocities (RV) ranging from 1.0 to 5 km/s (Figure 8). Our filtering options fits with those used by *Kim and Dreger* (2008) in their study (low-pass filter at T = 2 s). Four sites were dominated by either near-field 3D effect(s) that could not be modelled using a smooth 1D velocity profile (CARH, POMM, MNMC) and/or clipping (MIDA). We used a set of Green's function based on the 1D seismic model GIL7[77]. Green's functions (different from those used by Kim and Dreger, 2008) were designed to render both seismic wave arrivals and the static offsets for data collected in the very near field. The variance reduction for the various models is always larger than 60%. The total moment release estimates are between 9 and 14×10$^{24}$ dyne.cm (Figure 8). The slip pattern is stable for various rupture velocities except toward the southern end of the rupture where coverage is poor. We show that waveform fits are good for solutions with rupture velocities between 2 to 3 km/s (Figure 9). The same work detailed below has been completed using an asymmetric model[78] without improving the inversion variance reduction or changing the shape of slip distribution at depth (Supplementary Materials Figures B3, B4 and B5 are displaying these results). The computed slip distribution solutions (Figure 10) also fit well the spatial slip distribution inferred from the accelerometer records[53]. However, the top 5 percent of the slip values exceed (up to 125 cm) the maximum slip values (50 cm) constrained by other techniques (Table 2). However, the 95$^{th}$ percentile does not vary significantly with rupture velocity (Figure 8) with a maximum slip amplitude inferior to 67 cm. The peak slip area is close to the epicenter locations of the 1966 Parkfield mainshock and of the high-frequency 2004 Parkfield secondary event[79]. An area of low slip (10 cm or less) between the 1966 and the 2004 hypocenters is visible as for other models[53,80]. We expect that the slip to be poorly constrained near the southern end of the fault (epicentre area) where the GPS network is sparse at the time of the rupture. Bootstrap inversion tests (see in particular slip models MH and MV in Supplementary Materials) show that the slip amplitudes near the hypocenter are, in fact, constrained by the horizontal components. The slip model constrained by GPS compares visually well with other slip models based on strong motion data[52] computed with a different algorithm. The differences between strong motion model[53] computed using the same software are limited between −10 and +10 cm. This value fits well with standard deviation for slip proposed by *Page et al.*, (2009). We therefore assume that 10 cm as the 1$\sigma$ uncertainty on the slip, and we therefore are not able to detect significant difference between our model and the one proposed by *Kim and Dreger*, (2008) at the southern end of the fault. We then compare the slip distribution inferred from the inversion of our GPS waveforms (RV = 2.2 km/s, and RV = 2.6 km/s) with a slip model inferred from SAR interferometry (InSAR)[80]. We find discrepancies on the slip amplitude above the 5 km depth along the northern part of the Parkfield segment (Figures 10 and 11, as suggested for the first 24 hours by *Johnson*

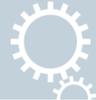

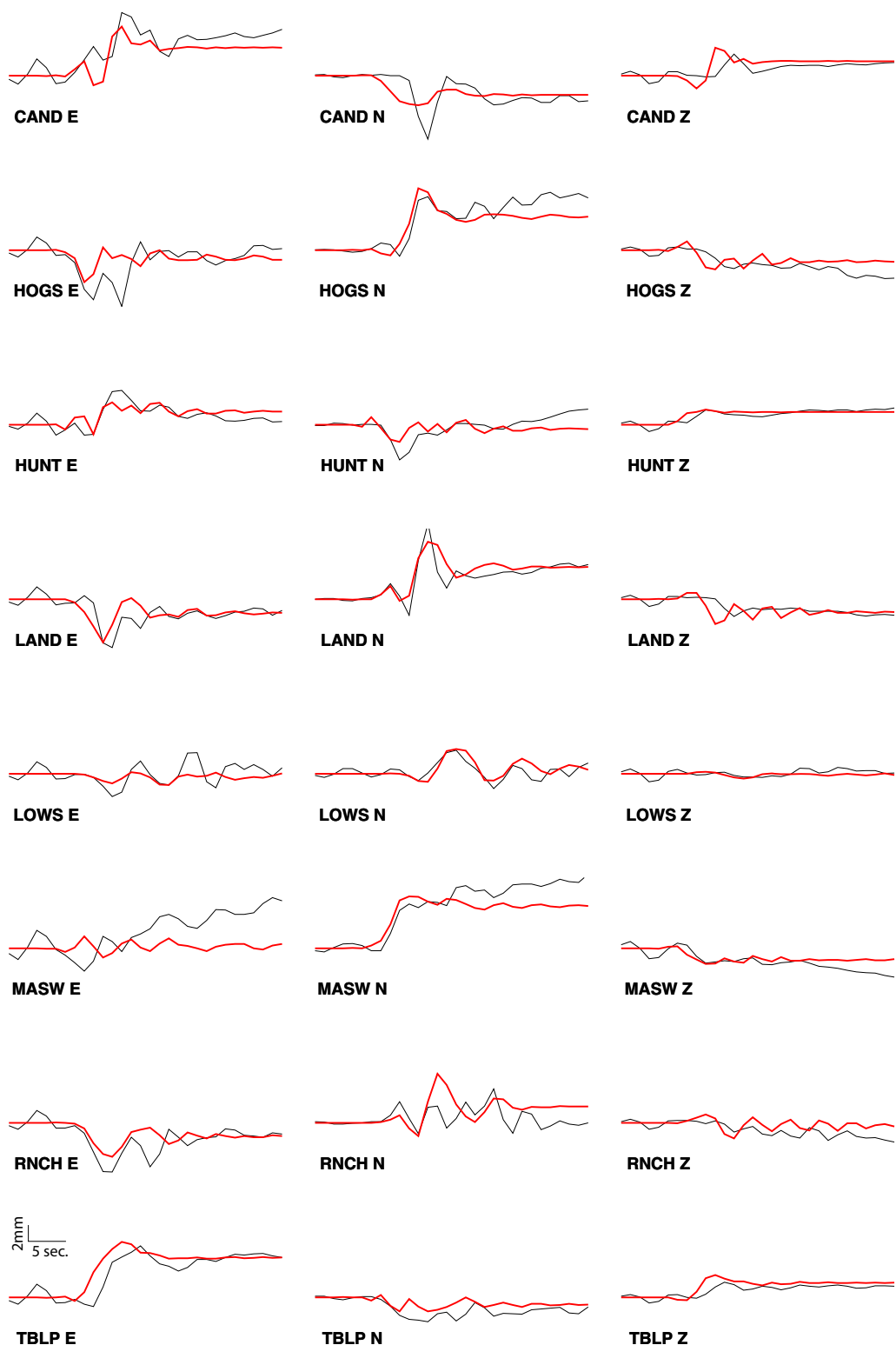

**Figure 9** | Fit between the GPS waveforms (black) and the model (red) for RV = 2.6 km/s.

*et al.*, (2006). As the resolution of InSAR is close to ±2 mm/yr for best configurations[62], and we showed that the near-surface slip must be well constrained (see section) near the surface if measurement points are densely distributed at the surface. We conclude this subsurface slip is real but not co-seismic. Other slip models based on daily GPS solutions confirm this hypothesis[65]. The concentration of microseismicity at shallow depth during the days following the mainshock also supports this hypothesis[81] (Figure 11).

## Discussion

The spatial distribution of the GPS network was not optimal at the time of the Parkfield earthquake. Today, the sensitivity of the continuous GPS network, after the deployment of the PBO GPS network would allow a better description of slip at all depth. The radiated energy from the first asperity is concentrated in higher frequency[53] which explains that 1 Hz GPS time-series could not constrain slip near the epicenter regardless of the good sensitivity of the GPS hori-zontal components. Along the northern part of the Parkfield segment, we retrieved a slip model at depth that is similar to those obtained from seismic records with a new set of Green's functions and/or different inversion algorithms. This implies that inversion input parameters do not drive the slip model characteristics. The mapping of slip at depth allows us concluding that the post-seismic slip locations coincides with areas where the aftershocks occurred during the 2 days following the mainshock at least along the northern section of the SAF at Parkfield (Figure 11). The comparison of our models with the model obtained from InSAR[80] confirm that post-seismic slip in the 0 to 5 km depth range initiated minutes after the end of the mainshock. This suggest the surface fractures mapped within the fault zone[82] were not co-seismic, but were instead created by post-seismic processes following the mainshock[83].

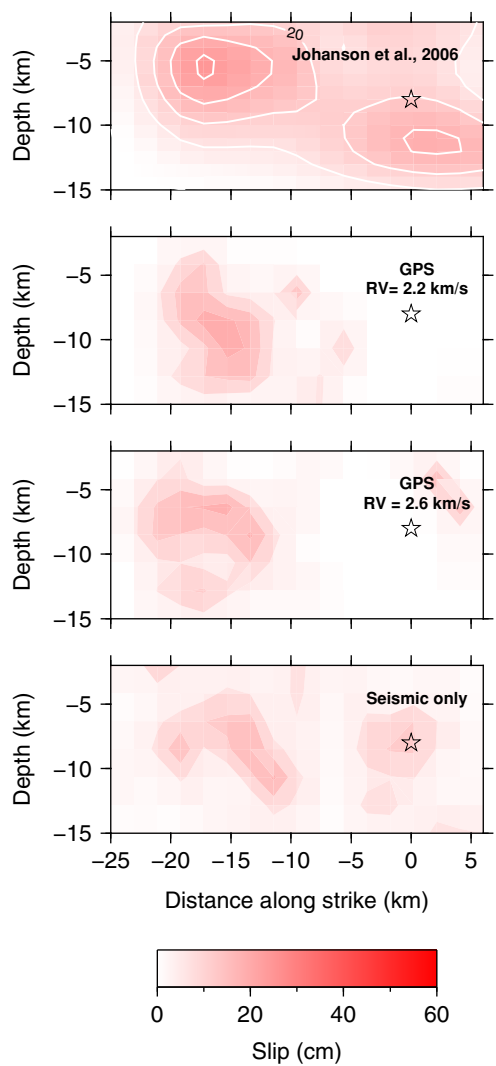

Figure 10 | **Slip distribution from various datasets.** We note that the GPS models are able to map the slip distribution as well as the seismic solutions. The slip to the southern tip of the fault is not well constrained likely because the GPS network was not deployed above the 2004 hypocenter. Models from Johanson et al., 2006 (white contour lines every 10 cm) and Kim and Dreger, 2008 are models inferred using InSAR and Strong-motion data only.

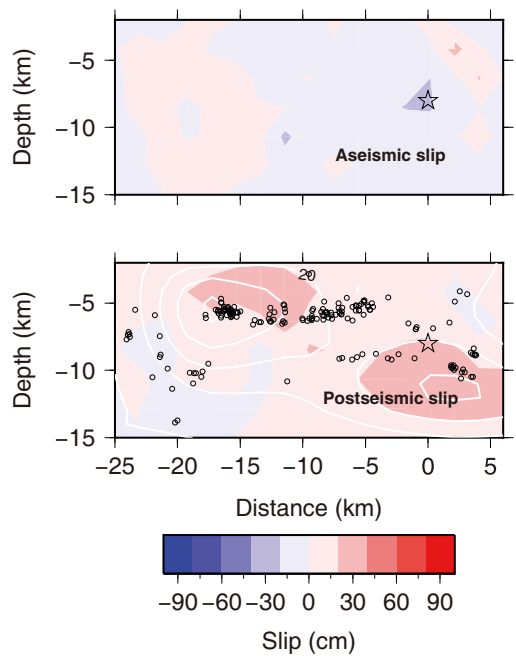

Figure 11 | **Aseismic (top) and Postseismic (bottom) slip distributions.** We subtracted the GPS dynamic solution (RV = 2.6 km/s) to the seismic and InSAR slip solution. We note that apart of the overestimation of slip in the area of the peak slip, all the late slip is concentrated between the surface and 5 km depth, or in an area surrounding the actual hypocenter of the source and in the zone between the 1966 and the 2004 hypocenter areas. We note that the area where post-seismic slip occurs coincides with the location of micro-earthquakes happening between the Parkfield mainshock and the second SAR acquisition ($\sim T_0 + 2\ days$). The InSAR model discussed in the text is indicated by using white line contours (a line every 10 cm).

## Methods

Beyond the challenge of what is an optimum network geometry for recovery of coseismic slip using only GPS data, there remains, the question of the depth sensitivity of GPS observations to slip at depth. To examine this we model rectangular dislocations of various depth extent for two GPS networks at Parkfield: the GPS network as it was in place in September 2004 and today's GPS network (including the newest PBO sites). The dislocation's slip amplitudes range from 10 to 50 cm (the maximum displacement) for 3 sources rupturing the entire length (25 km) of the SAF segment: from 0 to 5 km, 5 to 10 km and 10 to 15 km (Figure A1). We find that in 2004, the GPS network would not be able to detect displacement if the slip was smaller than 20 cm and shallower than 5 km (Figure A1). The addition of InSAR to the GPS data is therefore necessary to quantify accurately the early postseismic deformation that occurred near the surface. The addition of Plate Boundary Observatory (PBO) sites since 2004–2005 allowed increasing the number of short baselines, and decreased the minimal baseline length; two improvements allowing leading to better resolution of the shallow deformation near the fault.

Table 2 | Statistics of the models for various instrumentations. We note while the variance reductions (Var.) are similar the time span range from seconds to days. Rupture Velocities (R. V.) are indicated when relevant

| Instr. | Time span (s) (s) | Var. | R.V. (km/s) | Mean slip (cm) | ±1σ | Ref. |
| --- | --- | --- | --- | --- | --- | --- |
| InSAR | $\sim 1.5\ 10^5$ | 90% | N/A | 12.0 | 11.12 | *Johanson et al.*, 2006 |
| Accelerograph | 30 | 60% | 2.6 | 5.6 | 7.8 | *Kim and Dreger*, 2008 |

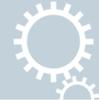

### Acknowledgments
Authors thank Pr. Chen Ji for his useful comments. The main author was supported by USGS through cooperative agreement 07HQAG0031 and by the University of California at Berkeley. The deployment of instrumentation was funded by the National Science Foundation under the Major Research Instrumentation program, Grant No. 9977823 with matching funds from the Berkeley Seismological Laboratory, the Department of Terrestrial Magnetism at Carnegie Institution of Washington, the IGPP at UC San Diego, U.S. Geological Survey at Menlo Park, Calif. and the Southern California Earthquake Center. Maintenance and operations are now supported by the USGS under Cooperative Agreement G10AC00093. The COSMOS VDC is supported by funds from the National Science Foundation (CMS-0201264). This study is a Berkeley Seismological Laboratory contribution. All Figures have been drawn with Generic Mapping Tools.

### Author contributions
N.H. initial idea, conception and writing; N.H. and D.D. wrote the text. A.K. completed the first set of inversion.

### Additional information
**Supplementary information** accompanies this paper at http://www.nature.com/scientificreports

**Competing financial interests:** The authors declare no competing financial interests.

**How to cite this article:** Houlié, N., Dreger, D. & Kim, A. GPS source solution of the 2004 Parkfield earthquake. *Sci. Rep.* **4**, 3646; DOI:10.1038/srep03646 (2014).

Supplementary Materials: "GPS source solution of the 2004 Parkfield earthquake"

By Houlié, N, D. Dreger, and A. Kim.

# 1. Figures

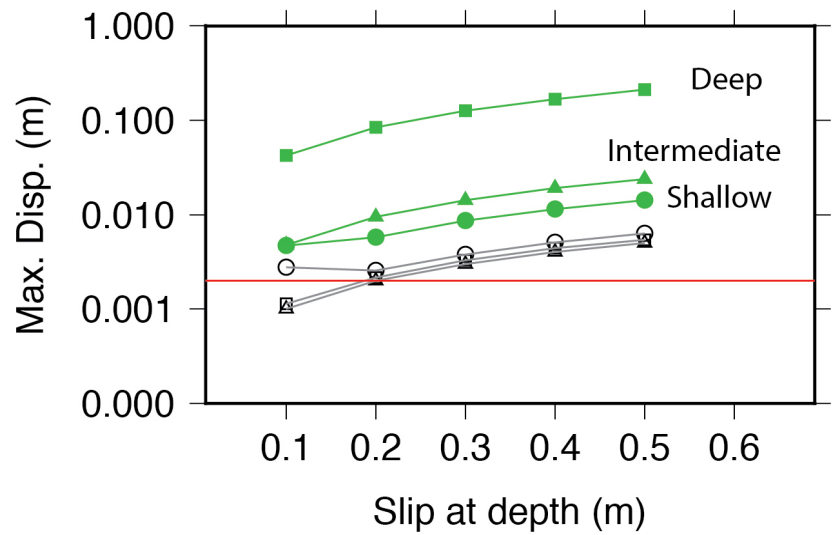

Figure A1: Maximum displacement potentially detected by sites located at Parkfield. The three models are rupturing the entire length of the segment (25km) for three ranges of depth (0-5km, 5-10km, 10-15km). They are respectively represented by squares, triangles and circles.

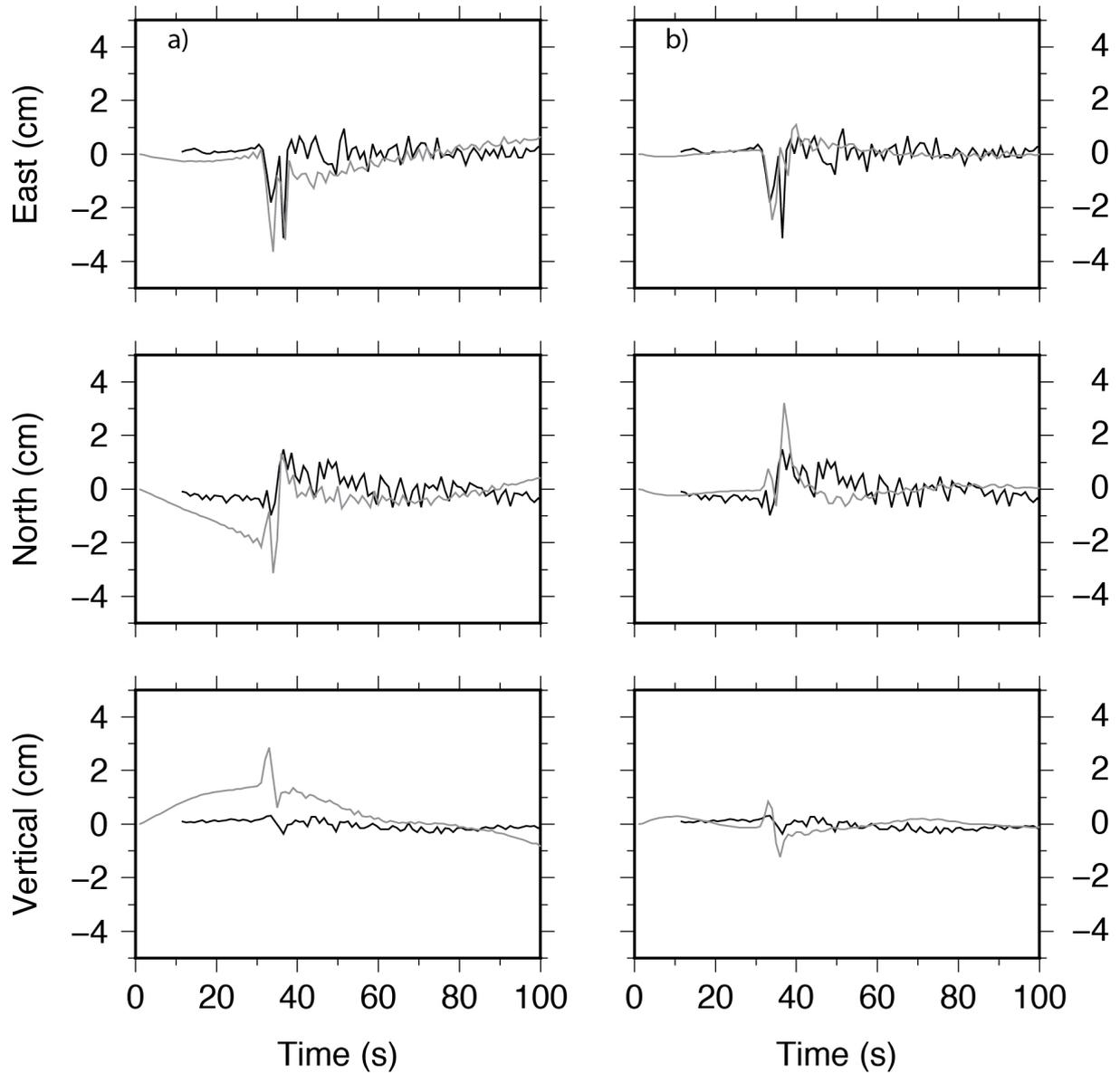

Figure A2: a) Comparison of the GPS times series (HOGS in black) with the strong motion sensor collocated (PHOB in gray). The accelerograms have been decimated at 1Hz and lowpass-filtered at 3s to match GPS waveforms characteristics. b) Same as a) with accelerograms bandpass filtered between 3 and 50s.

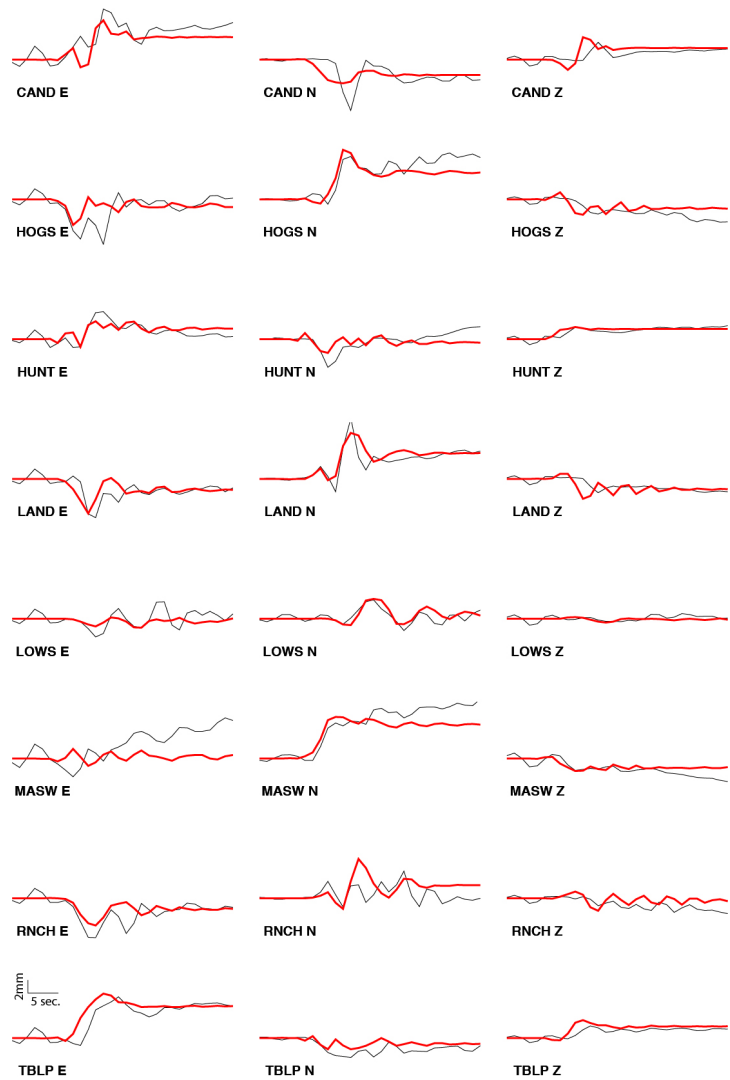

Figure A3: Same as Figure 9 for the asymmetric velocity model. GPS waveforms (black) and the model (red) are compared for RV=2.6km/s

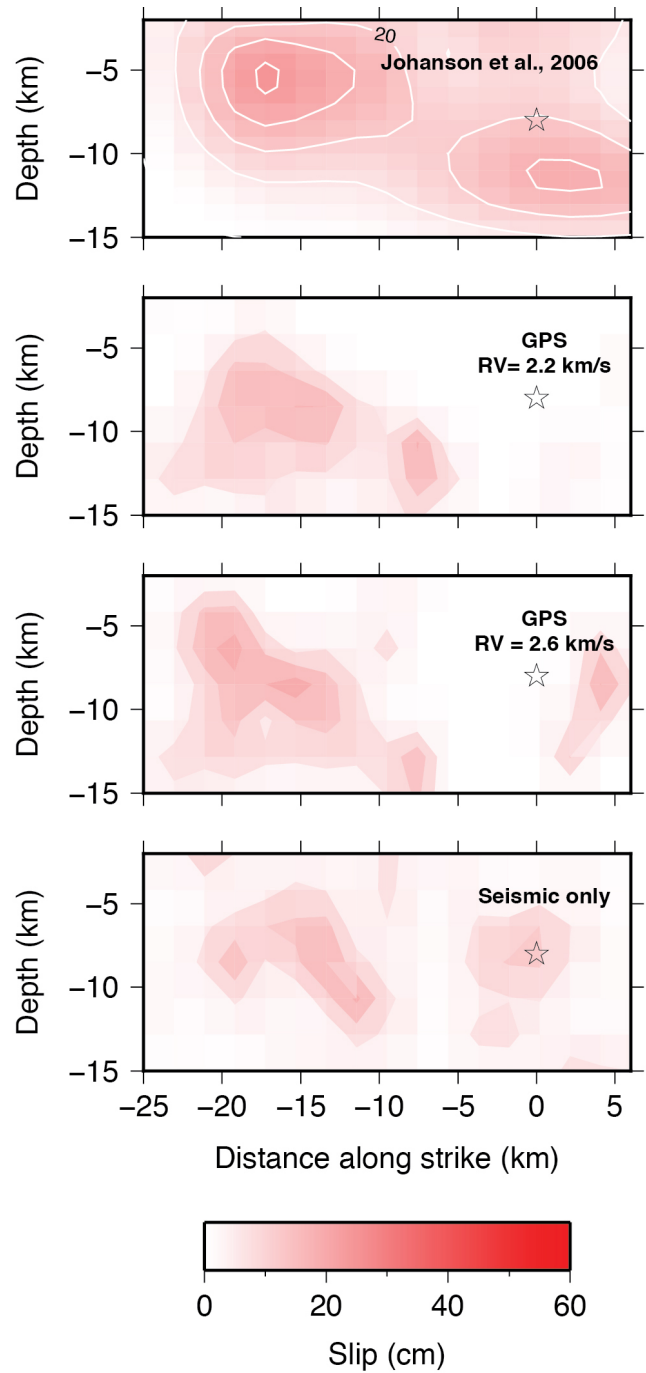

Figure A4: Same as Figure 10. Slip distribution from various dataset. The GPS solutions are computed using the asymmetric model used by Kim and Dreger (2008).

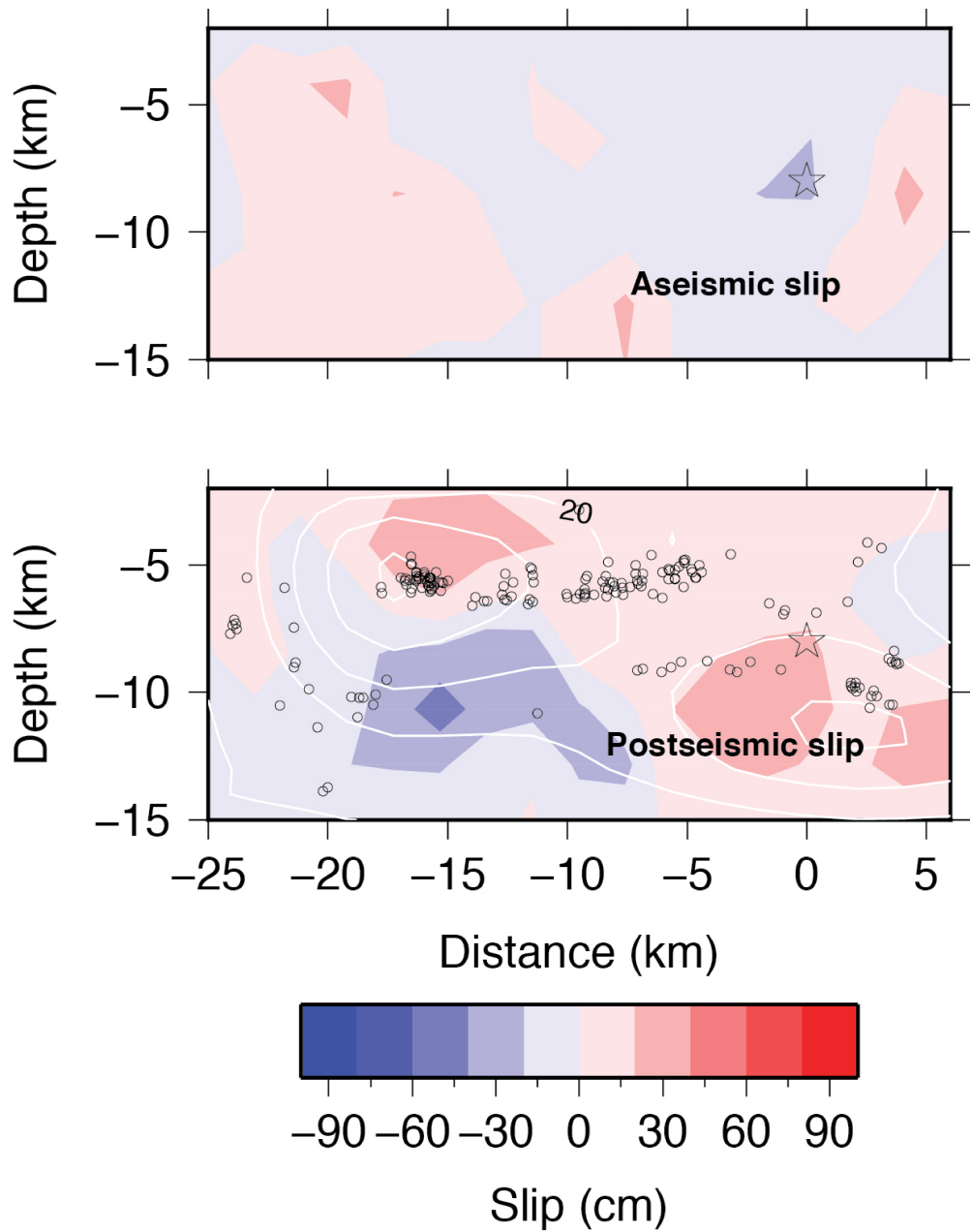

Figure A5: Same as Figure 11 for the asymmetric velocity model.

## 2. Bootstrap models

We display here the results of the bootstrap inversions completed. Parameters of each inversion are listed in Table S1.

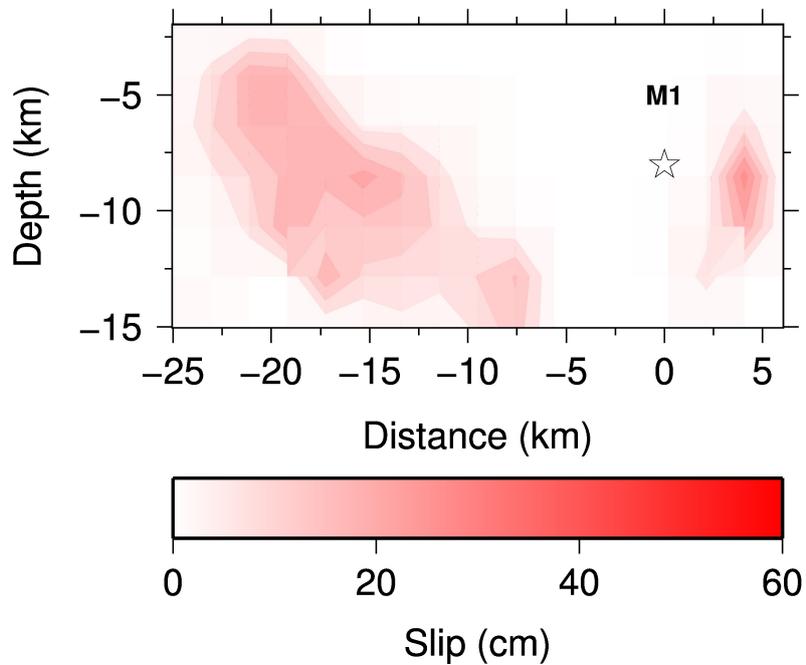

Figure B1: Slip distribution (model M1) if we exclude the data of the site CAND. All components are used.

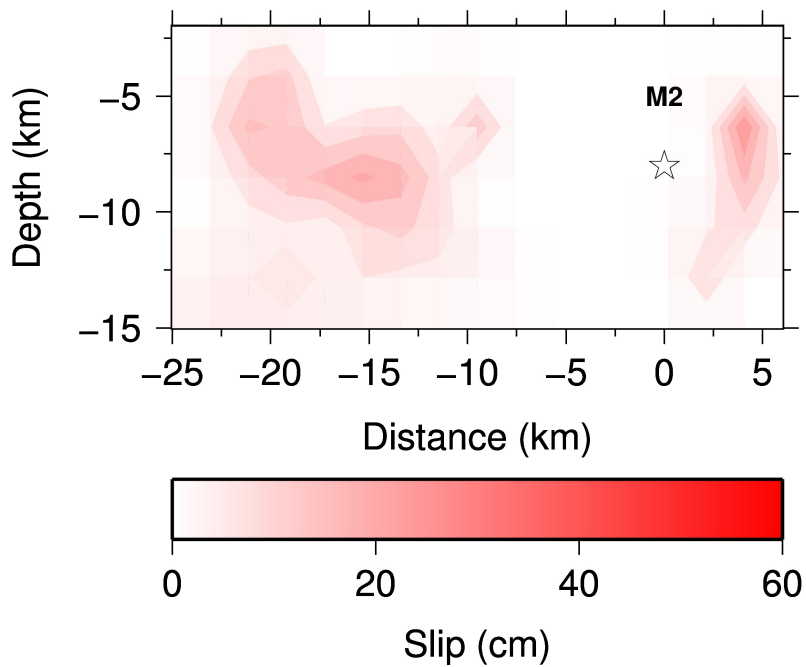

Figure B2: Slip distribution (model M2) if we exclude the data of the site HOGS. All components are used.

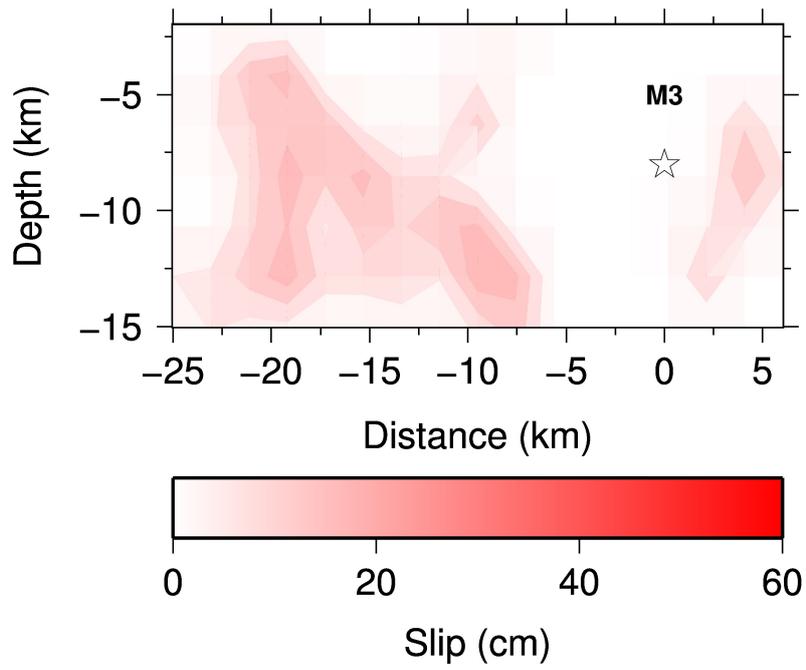

Figure B3: Slip distribution (model M3) if we exclude the data of the site HUNT. All components are used.

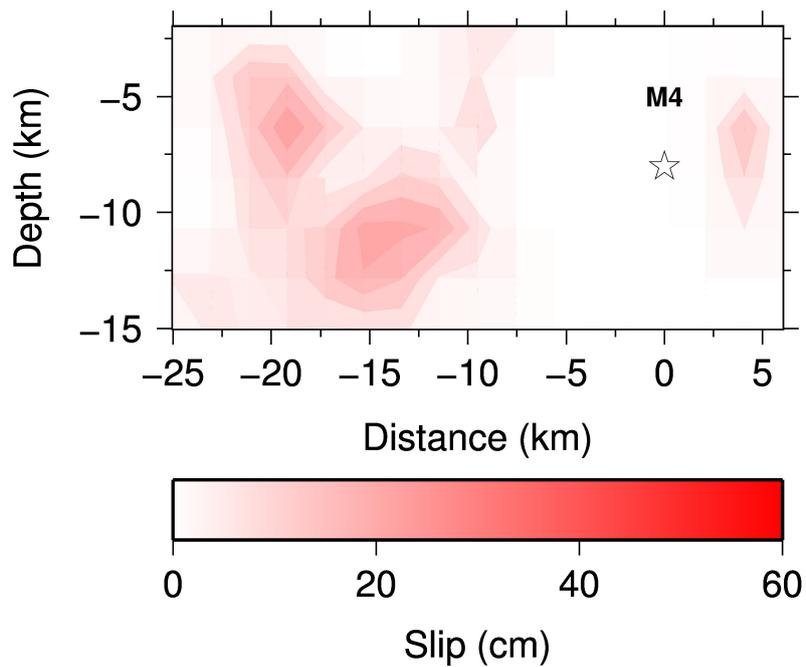

Figure B4: Slip distribution (model M4) if we exclude the data of the site LAND. All components are used.

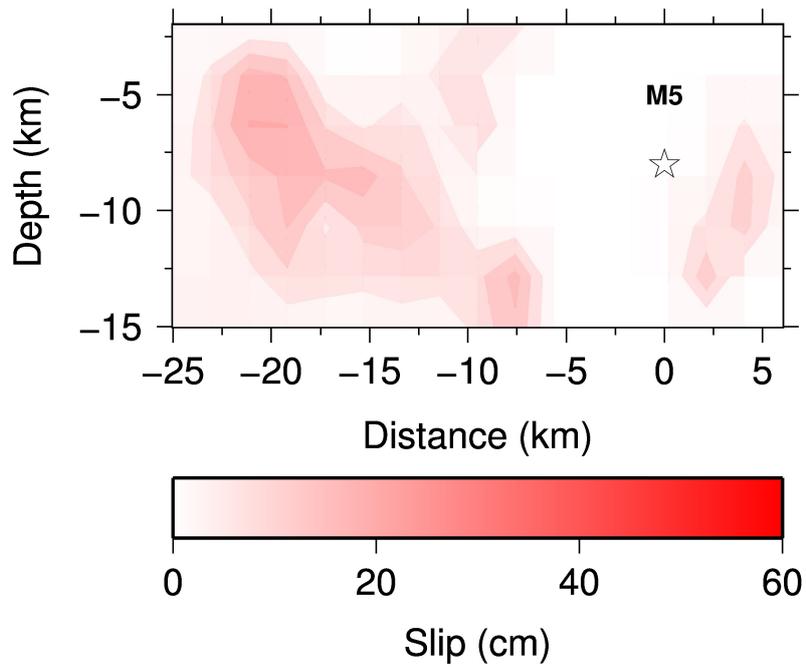

Figure B5: Slip distribution (model M5) if we exclude the data of the site LOWS. All components are used.

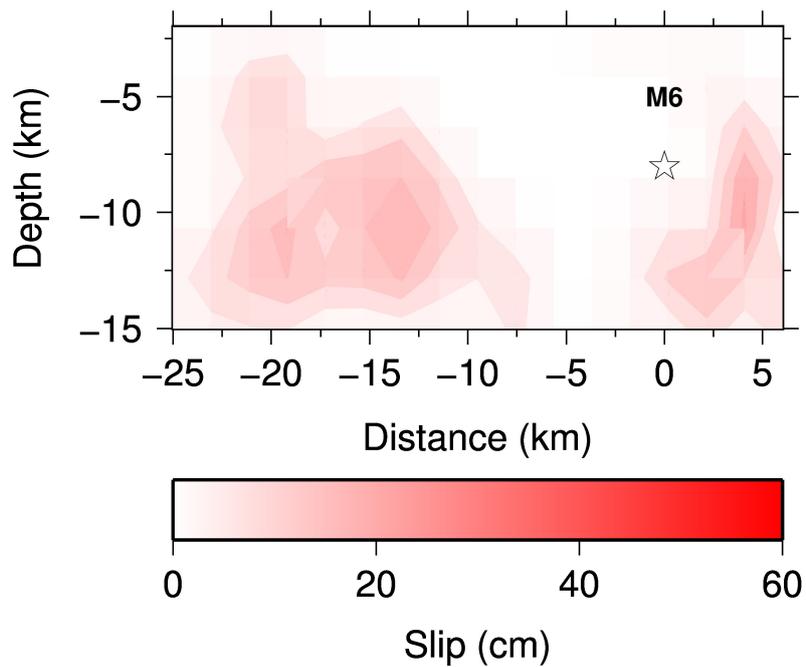

Figure B6: Slip distribution (model M6) if we exclude the data of the site MASW. All components are used.

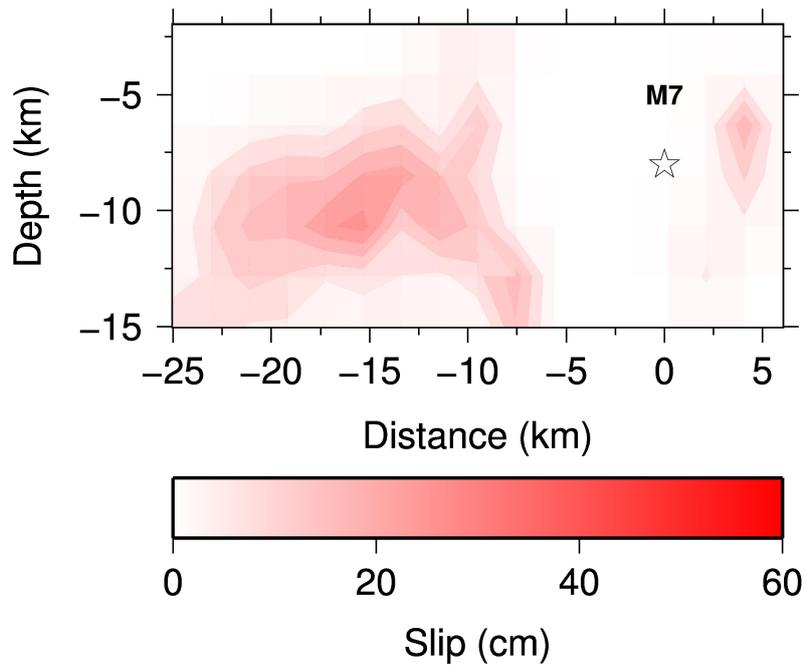

Figure B7: Slip distribution (model M7) if we exclude the data of the site RNCH. All components are used.

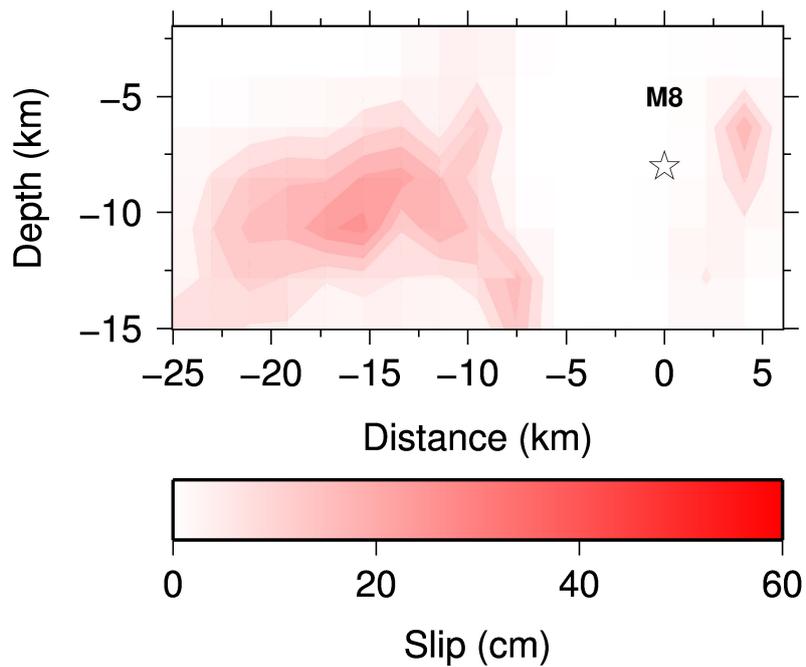

Figure B8: Slip distribution (model M8) if we exclude the data of the site TBLP. All components are used.

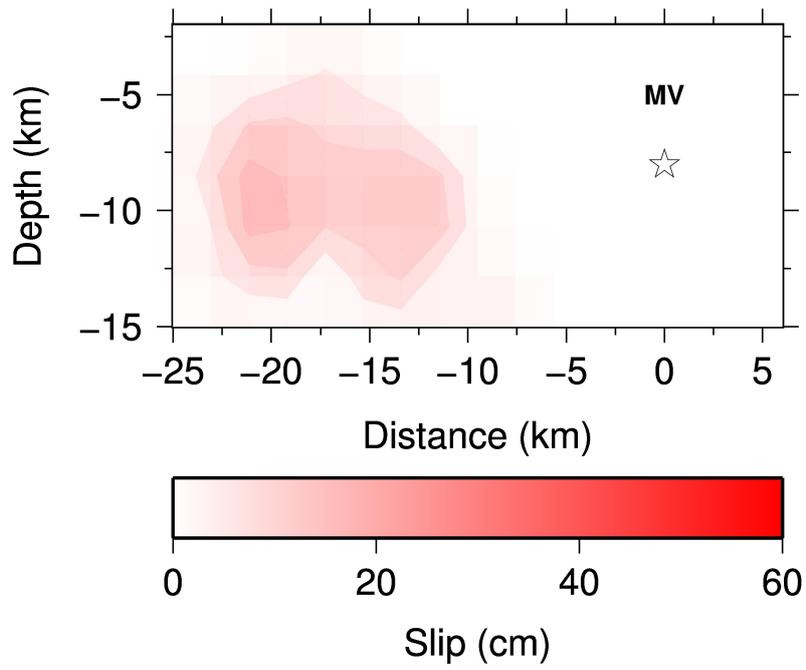

Figure B9: Slip distribution if we exclude horizontal components of the inversion.

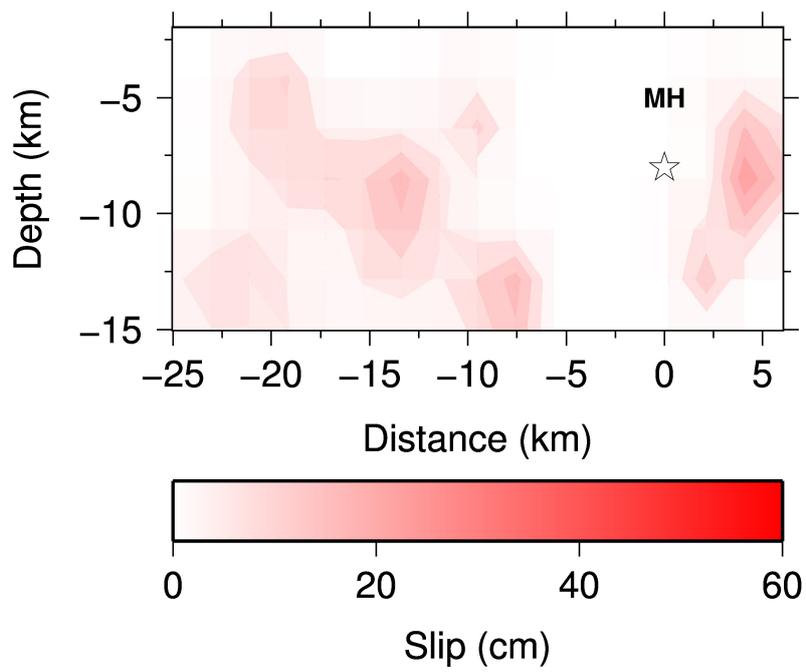

Figure B10: Slip distribution if we exclude vertical components of the inversion.

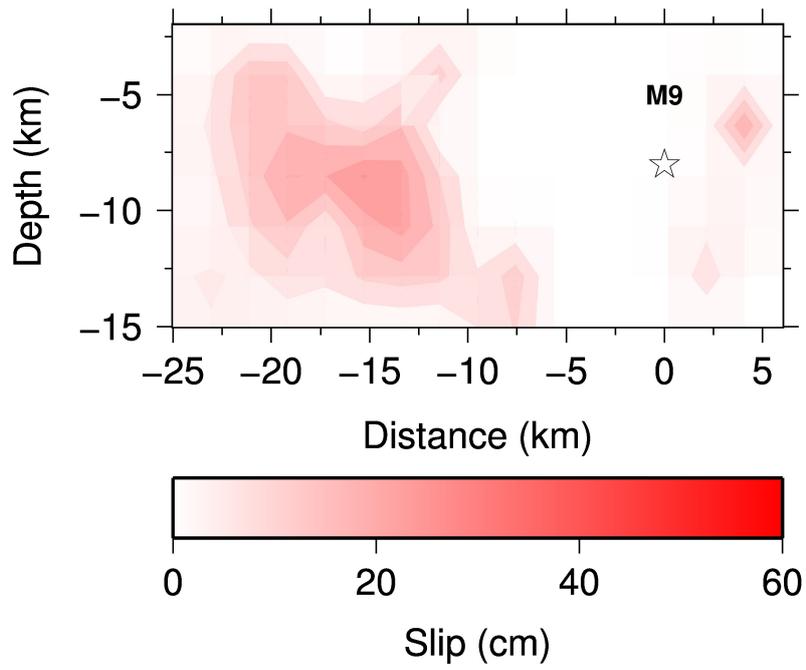

Figure B11: Slip distribution if we exclude vertical components of the inversion.

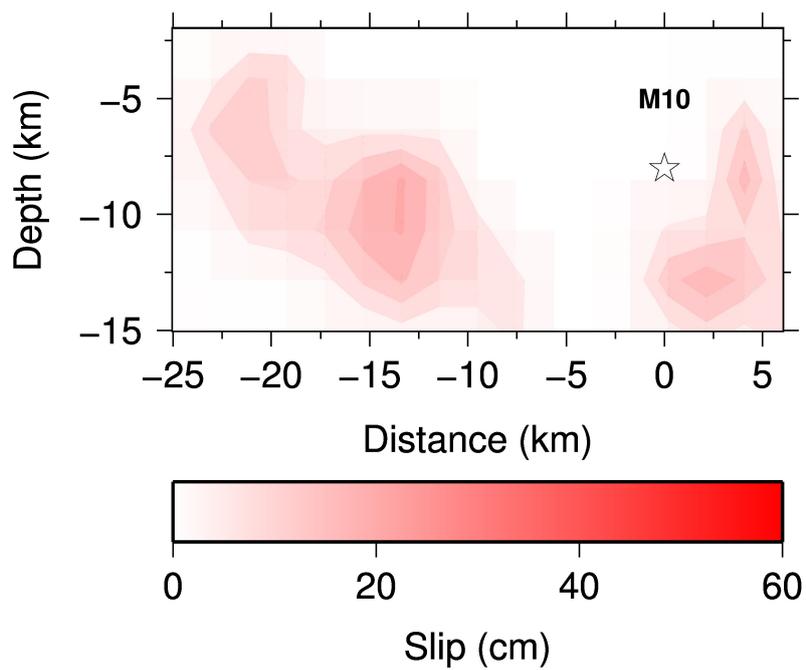

Figure B12: Slip distribution (model M1) if we exclude the data of the sites LOWS and TBLP. All components are used.

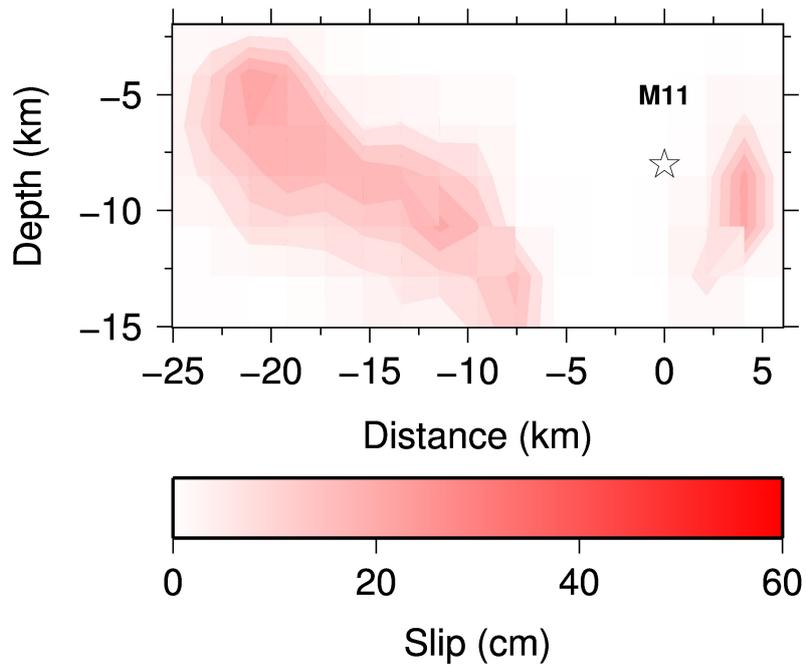

Figure B13: Slip distribution (model M1) if we exclude the data of the sites LOWS and CAND. All components are used.

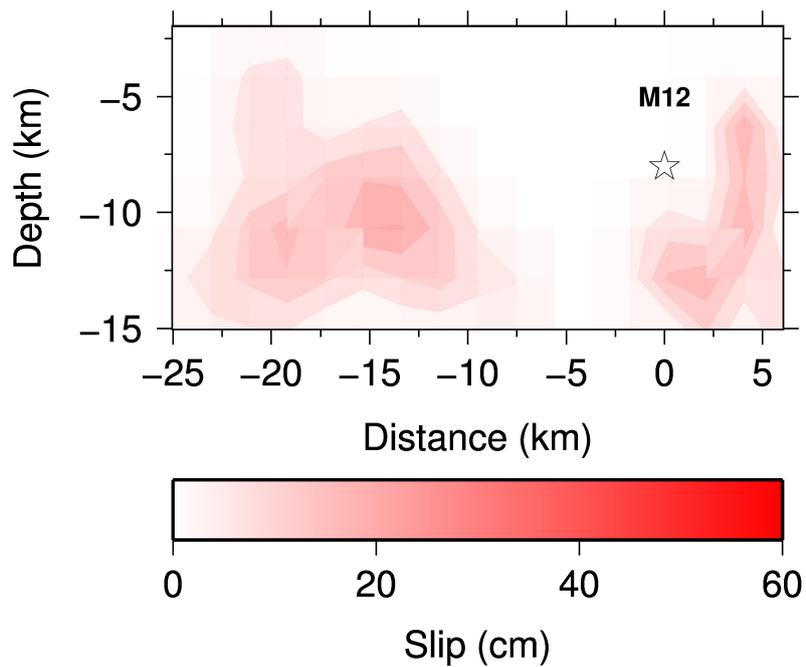

Figure B14: Slip distribution (model M1) if we exclude the data of the sites TBLP and MASW. All components are used.

| Figure | Model | Site(s) excluded | Number of components | Max slip (cm) |
|---|---|---|---|---|
| B1 | M1 | CAND | 21 | 62 |
| B2 | M2 | HOGS | 21 | 60 |
| B3 | M3 | HUNT | 21 | 57 |
| B4 | M4 | LAND | 21 | 61 |
| B5 | M5 | LOWS | 21 | 63 |
| B6 | M6 | MASW | 21 | 56 |
| B7 | M7 | RNCH | 21 | 66 |
| B8 | M8 | TBLP | 21 | 66 |
| B9 | MV | All horizontal components | 8 | 56 |
| B10 | MH | All vertical components | 16 | 63 |
| B11 | M9 | LOWS and TBLP | 18 | 68 |
| B12 | M10 | CAND and MASW | 18 | 58 |
| B13 | M11 | LOW and CAND | 18 | 59 |
| B14 | M12 | TBLP and MASW | 18 | 58 |

Table S1: Description of the models completed to control the resolution of slip at depth.

| RV (km/s) | Ave. Slip (cm) | Max. Slip (cm) | STD (cm) |
| --- | --- | --- | --- |
| 1.0 | 6.9 | 97.3 | 17.8 |
| 1.2 | 7.5 | 132.0 | 20.4 |
| 1.4 | 7.5 | 136.7 | 19.4 |
| 1.6 | 7.2 | 137.3 | 21.6 |
| 1.8 | 6.9 | 97.3 | 17.8 |
| 2.0 | 6.7 | 107.9 | 19.0 |
| 2.2 | 6.7 | 125.9 | 19.5 |
| 2.4 | 7.4 | 101.9 | 20.6 |
| 2.6 | 7.6 | 131.7 | 20.2 |
| 2.8 | 8.4 | 155.2 | 22.5 |
| 3.0 | 8.9 | 114.5 | 20.1 |
| 3.2 | 9.4 | 109.9 | 20.7 |
| 3.4 | 8.9 | 142.2 | 21.8 |
| 3.6 | 8.9 | 101.9 | 19.9 |
| 3.8 | 9.0 | 99.9 | 19.9 |
| 4.0 | 9.1 | 83.5 | 19.5 |
| 4.2 | 8.8 | 116.2 | 20.9 |
| 4.4 | 8.5 | 118.2 | 21.5 |
| 4.6 | 8.4 | 120.8 | 20.6 |
| 4.8 | 8.2 | 94.5 | 19.3 |
| 5.0 | 8.1 | 113.5 | 21.5 |

Table S2: Average, maximum slip and standard deviation for each rupture velocity (RV).